\documentclass[aps,prl,twocolumn,superscriptaddress,showpacs]{revtex4}
\usepackage{amsmath,amsthm,amsfonts}
\usepackage{epsf}
\usepackage{amssymb}
\usepackage{graphicx}
\usepackage{color}      

\newcommand{\ket}[1]{| #1 \rangle}
\newcommand{\bra}[1]{\langle #1 |}

\newcommand{\ex}[1]{\langle #1 \rangle}

\newcommand{\beq}{\begin{eqnarray}}
\newcommand{\eeq}{\end{eqnarray}}

\begin{document}
\title{Non-equilibrium Entanglement and Noise in Coupled Qubits}
\author{N.~Lambert}
\affiliation{The University of Tokyo, Department of Basic Science,
3-8-1 Komaba, Meguro-ku, Tokyo 153-8902, Japan}
\author{R.~Aguado}
\affiliation{Departamento de Teor\'{\i}a de la Materia Condensada,
Instituto de Ciencia de Materiales de Madrid,
CSIC, Cantoblanco 28049, Madrid, Spain}
\author{T.~Brandes}
\affiliation{The University of Manchester, School of Physics and Astronomy, P.O. Box 88,
Manchester, M60 1QD, U.K.}
\bibliographystyle{apsrev}
\begin{abstract}
We study charge entanglement in two Coulomb-coupled double quantum
dots in thermal equilibrium and under stationary non-equilibrium
transport conditions. In the transport regime, the entanglement
exhibits a clear switching threshold and various limits due to
suppression of tunneling by Quantum Zeno localisation or by an
interaction induced energy gap. We also calculate quantum noise
spectra and discuss the inter-dot current correlation as an
indicator of the entanglement in transport experiments.
\end{abstract}
\maketitle


Precise engineering and preparation of entangled states forms the
backbone of many quantum information schemes \cite{Chuang}. The
complete control of interactions between two or more parties is a
sangraal that is not without cost. For example, in superconducting
nano-circuits \cite{NPT99} there has been much success in devising
schemes for {\em tunable} capacitative couplings \cite{Averin03},
but thermal fluctuations, background noise and limited control over
{\em natural} interactions  must be dealt with and overcome in
increasingly imaginative ways.

In this Letter, we take a slightly different point of view and ask
for the degree of entanglement between two parallel, interacting
electronic conductors under the `un-favourable' condition of
stationary currents passing through both of them. As this is
clearly a mixed-state situation, we specifically consider a
non-equilibrium version of the concurrence as entanglement measure
for an electron charge double qubit (DQ), realised in
Coulomb-coupled double quantum dots \cite{Hayetal03,Bra05_a} that
are strongly coupled to external electron reservoirs at high
voltage bias. We compare this to the same closed device in
equilibrium with a heat bath, and our findings suggest that such
an approach, although probably not directly relevant for quantum
information purposes, sheds a new light on the relation between
entanglement and the electronic transport process itself. In
particular, effects like suppression of non-resonant tunneling and
the quantum Zeno effect (QZE) have a direct impact on the
entanglement, to which we also establish a further link by
calculating the non-equilibrium quantum shot-noise tensor whose
off-diagonal elements, as a function of the system parameters,
show a behavior very similar to the concurrence.

\begin{figure}[t]
\includegraphics[width=0.6\columnwidth]{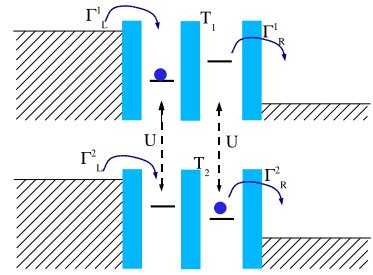}
\caption{Transport double qubit model: left-right charge states in
two Coulomb-coupled double quantum dots with one additional
electron each and  `on-site' ($LL$, $RR$) interaction $U$,
coherent tunnel couplings $T_1$ and $T_2$ and electron reservoir
tunnel rates $\Gamma^{1/2}_{L/R}$.} \label{fig1}
\end{figure}

{\em  Model.--} For the sake of clarity, we  define the double
qubit by `left' and `right' orbital charge states
$|\alpha_i\rangle$, $\alpha=L,R$ of one additional electron on top
of  the many-body  ground state $|0_i\rangle$  (limit of  {\em
intra}dot Coulomb interaction $U_{\rm in}\to \infty$) of two
double quantum dots $i=1,2$   which are coupled by a single matrix
element $U$ for  {\em inter}-dot 'same site' interactions
(left-left and right-right), cf. Fig. (\ref{fig1}). Tunnelling of
electrons occurs only within but not between the qubits due to
coupling $T_i$ in each double dot.  Using projectors onto these
orbital states, ($\hat{n}_L^{(i)}=\ket{L}\bra{L}_i$,
$\hat{n}_{LR}^{(i)}=\ket{L}\bra{R}_i$,...), the total Hamiltonian
is
\begin{eqnarray}\label{hamiltonian}
 {\mathcal H}_0&=&\sum_{i=1,2} \left( \varepsilon_i (\hat{n}_{L}^{(i)}-\hat{n}_{R}^{(i)}) +
T_i (\hat{n}_{LR}^{(i)} + \hat{n}_{RL}^{(i)}) \right) \nonumber\\&+&
\frac{U}{2}\left(\hat{n}_{L}^{(1)}\hat{n}_{L}^{(2)}+\hat{n}_{R}^{(1)}\hat{n}_{R}^{(2)}
\right).
\end{eqnarray}
The electron spin label is suppressed here and in the following,
as only charge states (acting as pseudo-spin) play a role.
This description has turned out to be useful for modelling
charge-related properties such as decoherence and noise in
individual double quantum dots \cite{AB04,Bra05_a}.

We  `open' the DQ by coupling  it to four external electron reservoirs,
${\mathcal H}={\mathcal H}_0 + {\mathcal H}_T+ {\mathcal H}_{\rm
res}$,
with ${\cal H}_{\rm res}=\sum_{i=1,2}\sum_{\alpha\in
L,R}\sum_{ki\alpha}\epsilon_{k i \alpha}c_{k i
\alpha}^{\dagger}c_{k i {\alpha}}^{\phantom{\dagger}}$ ($\alpha
=L/R$ refers to left and right reservoirs for qubit number $i$,
$i=1,2$)
and ${\cal H}_T=\sum_{i=1,2}\sum_{\alpha\in L,R}\sum_{k}
(V_k^{\alpha i} c_{k i {
\alpha}}^{\dagger}\hat{s}_{\alpha}^i+H.c.)$, with Hubbard
operators $\hat{s}_{\alpha}^i=|0_i \rangle \langle \alpha_i|$
that couple the qubits to the continuum.

\begin{figure}[t]
\begin{center}
 \includegraphics[width=0.48\columnwidth]{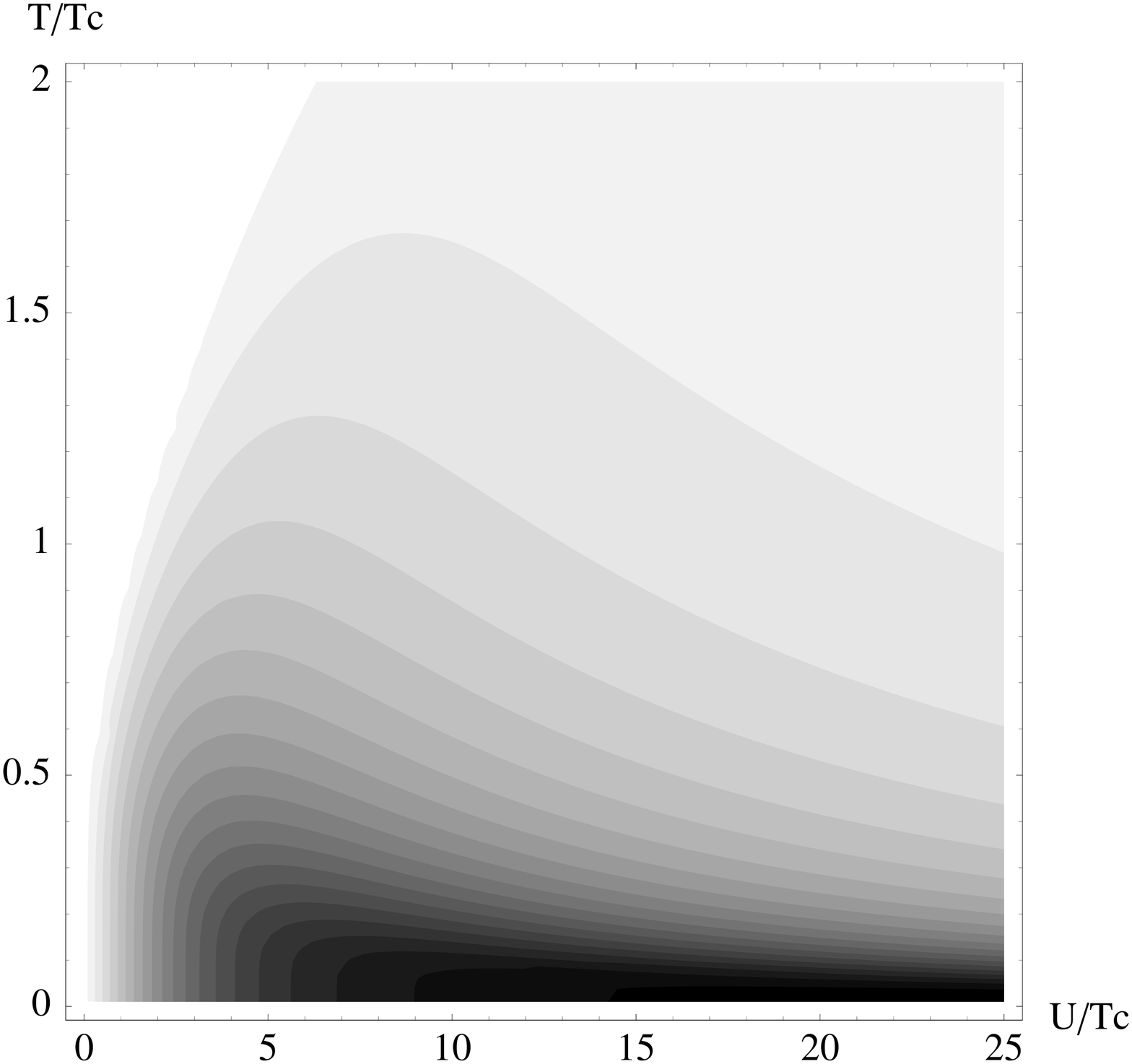}
\includegraphics[width=0.48\columnwidth]{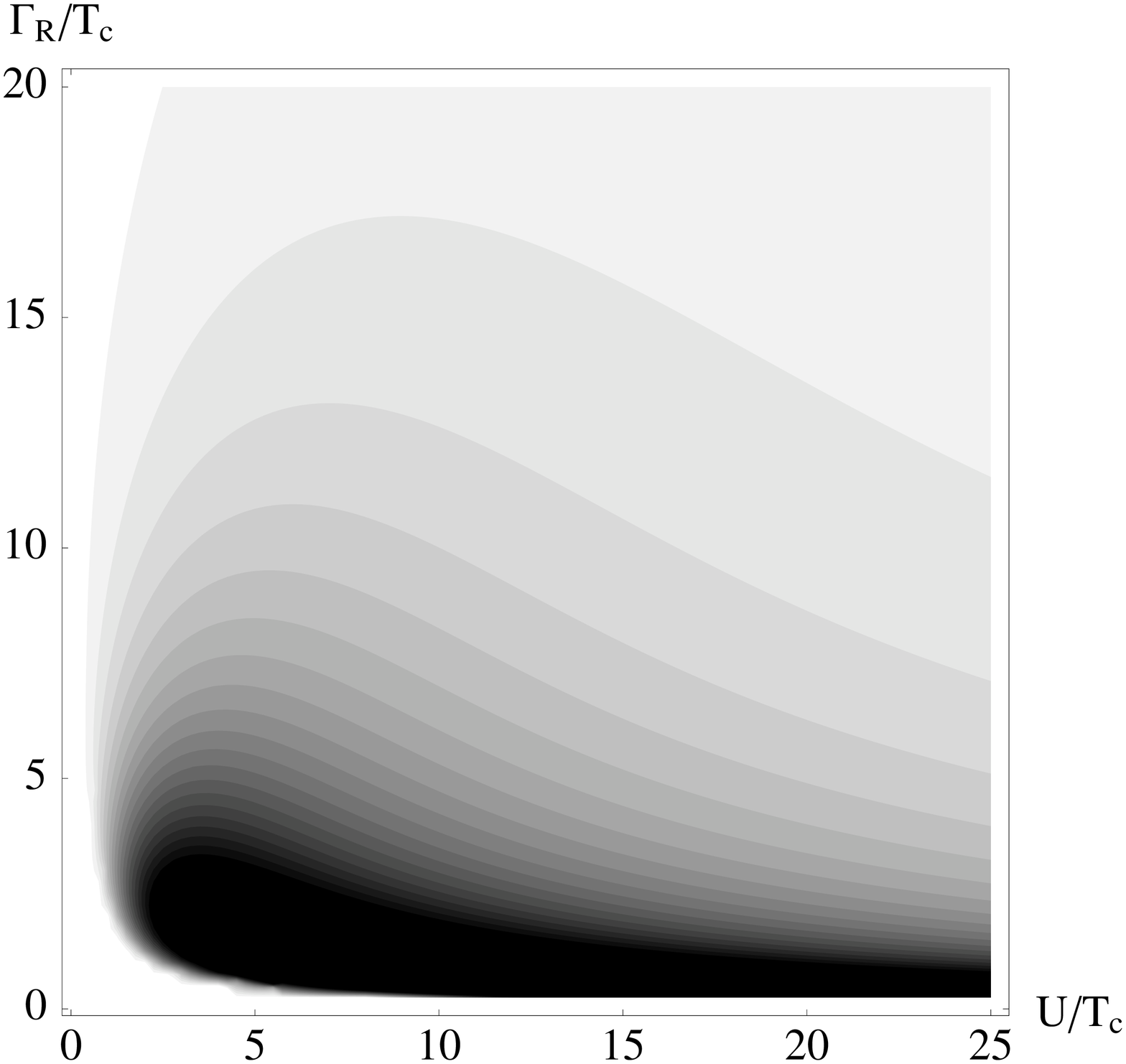}

\caption{\textbf{Left:} Grey-scale plot of double qubit equilibrium
concurrence as a function of interaction $U/T_c$ and temperature
$T/T_c$ (white is zero, black is maximum $C=1$ ).  In all of the
following results both devices have identical parameters
$\Gamma_L^{(i)}\equiv \Gamma_L$, $\Gamma_R^{(i)}\equiv \Gamma_R$, $T_i\equiv T_c$, for $i=1,2$.  
\textbf{Right:}  Concurrence of  non-equilibrium double qubit as a
function of interaction $U/T_c$ and reservoir tunnel rate $\Gamma_R$
for large $\Gamma_L/T_c=50$. Zero entanglement occurs below a
threshold $\propto 1/\Gamma_R$ in the weak tunnelling regime, and
for very strong tunnelling $\Gamma_R \gg T_c$ due to
Zeno-localisation, cf. text.
 (Black is the maximum $C=0.3$).
} \label{fig_eq_noneq}
\end{center}
\end{figure}

{\em Equilibrium entanglement.--} If the DQ is disconnected from the
reservoirs (${\mathcal H}_T=0$) but in contact with a heat bath at
temperature $T=1/\beta$, the equilibrium entanglement between qubit
1 and 2 is easily obtained from the concurrence \cite{Woo98} (a
well-known entanglement measure of mixed states of two qubits)
$C(\beta)$ for the canonical ensemble state $\rho(\beta)=e^{-\beta
{\mathcal H}_0}/Z$, $Z={\rm Tr} e^{-\beta {\mathcal H}_0}$. The
eigenvectors of ${\mathcal H}_0$ correspond to eigenvalues $E_0=0$,
$E_1=U$, and $E_{\pm}=(U\pm \sqrt{16T_c^2+U^2})/2$ and are expressed
in the basis of singlet and triplet states,
$S_0=1/\sqrt{2}(|L_1R_2\rangle-|R_1L_2\rangle)$,
$T_+=|L_1L_2\rangle$, $T_-=\ket{R_1R_2}$,
$T_0=1/\sqrt{2}(|L_1R_2\rangle+|R_1L_2\rangle)$. For simplicity we
restrict ourselves to the unbiased, symmetric case
$\varepsilon_i=0$, $T_i=T_c (i=1,2)$.  It turns out that the
equilibrium case already exhibits some interesting features, cf.
Fig.(\ref{fig_eq_noneq}). At any finite temperature $T$, the
entanglement is zero below a certain threshold value of the
interaction $U$ where the state $\rho(\beta)$ is too mixed in order to be
entangled which is, e.g., in analogy with the corresponding
transition in the (abstract) example of the Werner state
\cite{Wer89}. Furthermore, the concurrence shows a non-monotonic
behaviour as a function of $U$ at fixed $T$, with an entanglement
maximum at an optimal $U$-value.

{\em Stationary transport.--} The limit $t\to \infty$ in the
dynamical evolution of the reduced DQ density operator $\rho$
defines a stationary  non-equilibrium state $\rho_{\infty}$ which
usually is much more difficult to determine than in the equilibrium
case. Transport properties of models like Eq.(\ref{hamiltonian}) can
be analysed by using various non-equilibrium techniques. Here, we
consider a specific limit of infinite source-drain bias in order to
obtain quasi-analytic results from a generalised Master equation,
$\dot{\rho} =
{L}[\rho]$. The super-operator $L$ is
parametrized by the Markovian DQ-lead tunnel rates
$\Gamma_{\alpha}^i\equiv 2\pi\sum_{k} |V_k^{\alpha
i}|\delta(\varepsilon- \epsilon_{k i {\alpha}})$ (of which the
energy-dependence is neglected), and the DQ parameters 
$\varepsilon_i=0$, $T_i=T_c (i=1,2)$.  Analytical expressions for
the stationary solution of the 25 coupled equations of motion
(EOM) can then be found in an approximation where the broadening
due to tunnelling of the DQ levels is neglected, which for
$\varepsilon_i=0$, however, is only a very crude approximation.

One obtains
better results for the stationary currents $\langle
I_i\rangle_\infty$  by second order perturbation theory in the
intra-dot tunnel couplings $T_i$, which clearly show
a tunnel-broadened resonance 
\begin{eqnarray}\label{curr}
\langle I_i\rangle_\infty =-e
\frac{\Gamma_R^iT_i^2}{(\Gamma_R^{i}/2)^2+U^2}
\end{eqnarray}
($-e$ is the electron charge). In this limit, the
resonance is determined by the energy gap $U$ between the localised
eigenstates of the DQ: at large $U$, the triplet $T_+=|L_1L_2\rangle$
becomes populated (note that this state is always available because
of the infinite voltage approximation).
The energy gap to any other state involving delocalized electrons (e.g, the triplet $T_0$ or the singlet $S_0$)
then suppresses the elastic current.
In analogy to single charge qubits, where the energy gap is given by the internal bias $\varepsilon$,
we expect this 
suppression to be lifted in the presence of inelastic processes
\cite{Fujetal98}.

Furthermore, as a function of the coupling $\Gamma_R^i$ to the
drain, the current first increases and then becomes smaller again.
With the drains acting as broadband measuring devices (electron on
right side or not), strong couplings $\Gamma_R^i \to \infty$
completely freeze the  charges on the left sides which is a
`transport version' example \cite{CBLC04} of the QZE.
Alternatively, this localisation can be interpreted as an infinite
level broadening and the corresponding  suppression of the local
spectral density due to the decay to the drain.  Finally, the
behaviour
of the current, cf. Eq.(\ref{curr}), 
follows the occupation of the entangled singlet state $S_0$ as
illustrated in the inset of Fig.(\ref{fig_noise_spectrum}).  The
main current contribution 
therefore stems from two-particle tunneling events, which in turn
motivates our later comparison of the concurrence with the current
fluctuations.

{\em Non-equilibrium entanglement.--} We now define the
non-equilibrium entanglement via the concurrence $C$ of the
stationary state  $\hat{P}\rho_{\infty}$, where $\hat{P}$ is the
projection onto doubly occupied states including proper
normalisation;  i.e. we calculate the concurrence when both double
dots have a single electron in them and there are thus two two-state
systems to be entangled.
The projection $\hat{P}$ corresponds to taking the limit $\Gamma_L^i\to
\infty$ where both qubits are always occupied with one single
electron. For example, for $U=0$ and $\Gamma_L\to \infty$, the stationary state of a single charge qubit is described by the
(Bloch) vector of pseudo-spin Pauli matrices ($\varepsilon\equiv\varepsilon_L-\varepsilon_R$)
\begin{eqnarray}
\label{bloch}
\langle \vec{\sigma} \rangle =
\left(\frac{2T_c\varepsilon}{\mathcal{N}},
\frac{\Gamma_RT_c}{\mathcal{N}},\frac{\Gamma_R^2/4+\varepsilon^2}{\mathcal{N}}\right),
\end{eqnarray}
with $\mathcal{N}\equiv \Gamma_R^2/4+\varepsilon^2+ 2T_c^2$ and
in the $L$-$R$ basis where $\sigma_z\equiv |L\rangle \langle L| -|R\rangle \langle R|$ etc.
For $U\ne 0$, we  numerically checked  that $ \hat{P}\rho_{\infty}=\lim_{\Gamma_L^i\to\infty}\rho_{\infty}$ which means
that the two-qubit concurrence $C$ defined in this way does no longer depend on the left tunnel rates. This is
a good description of  non-equilibrium entanglement in a real system as long as
$\Gamma_L^i\gg$max$(U,\Gamma_R^i,T_c,\varepsilon)$.

As $C$ is zero to second order in $T_c$, we use numerical results, cf. Fig.(\ref{fig_eq_noneq}),
which shows an intriguing behaviour  of the concurrence as a
function of $U$ and the tunnel rate
$\Gamma_{R}^i\equiv\Gamma_R$.
We find a switching
threshold in that below an interaction strength $U\sim 2T_c^2/\Gamma_R$ the entanglement is
zero:
for small $\Gamma_R$, the stationary currents become very small, cf.
Eq. (\ref{curr}), and thus strong interactions are required in order
to entangle the dots. The DQ state becomes strongly mixed for
$\Gamma_R\to 0$ (note that we have not taken into account any
additional, internal relaxation processes here); its zero
entanglement along  the axis $\Gamma_R= 0$ is in fact a continuation of
the point $U=0$ where the states of both qubits are located at the origins of
their Bloch spheres, cf. Eq.~(\ref{bloch}).

On the other hand, for very large $\Gamma_R$ one runs again into
the QZE with electrons becoming trapped on the left ($\langle
\sigma_z\rangle\to 1$, cf. Eq.~(\ref{bloch})), and
$\hat{P}\rho_{\infty}$ approaching the (pure) localised state
$|L_1L_2\rangle$ which has zero entanglement. Finally, an increase
from small to larger $\Gamma_R$ at fixed $U$ yields the
re-entrance behaviour visible in the `teardrop'-shaped region of
large entanglement in Fig. ({\ref{fig_eq_noneq}).

\begin{figure}[t]
\includegraphics[width=1\columnwidth]{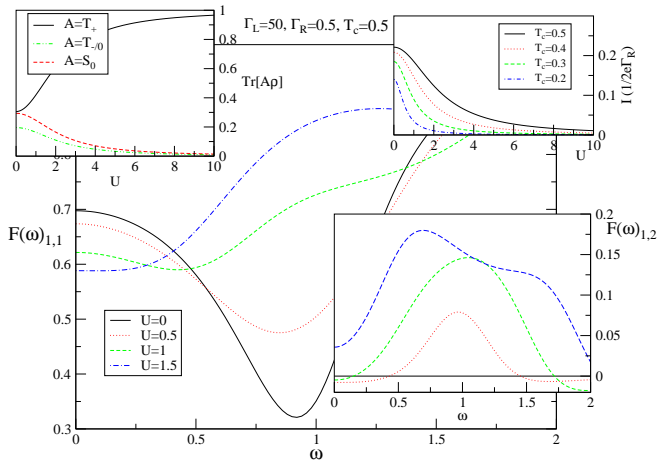}
\caption{\textbf{Main:}  Diagonal noise spectrum
$F(\omega)_{1,1}=S(\omega)_{1,1}/2eI$.
The resonance  at $\omega=2T_c$ splits into new
resonance points at the Bohr Frequencies $\lambda_{\pm}=E_1-E_{\pm}=1/2(U\mp\sqrt{16T_c^2 + U^2})$.
\textbf{Top right inset: } Stationary current $I_{\text{stat}}$. 
\textbf{Top left inset:}  Occupation of several singlet and triplet states.
\textbf{Bottom Inset:}  The cross-correlation frequency spectrum
$F(\omega)_{1,2}=S(\omega)_{1,2}/2eI$. 
Again resonance points manifest, however the correlation is always zero
for $U=0$ and can assume negative values for $U\ne 0$.} \label{fig_noise_spectrum}
\end{figure}

{\em Non-equilibrium  noise: formalism.--} Turning now to our
description of non-equilibrium shot-noise and its relation to
entanglement, the stationary  state  $\rho_\infty$ on its own is
not sufficient in order to describe intrinsic properties of the
DQ: for example, only limited information on the spectrum can be
obtained from stationary quantities like the current. In contrast,
the shot-noise spectrum exhibits resonances at the transition
frequencies of the system and contains furthermore useful
information on its relaxation and dephasing properties
\cite{Debetal03,BN03,KWS03,AB04}. We will now also  show an
emergent resemblance in the behaviour of the current {\em cross
noise} and the non-equilibrium concurrence as a function of the
system parameters.

In general, the finite-frequency noise has contributions from
particle currents as well as contributions from displacement
currents \cite{BB00,AB04}. In our case
($\Gamma_L^i\gg\Gamma_R^i$), however, it is a good approximation
to consider only particle currents.
Our starting point is the generating function
\begin{eqnarray}
 \hat{G}(s_1,...,s_m,t)=\sum_{n_1,...n_m=0}^{\infty} s_1^{n_1}\cdot\cdot\cdot s_m^{n_m}{\rho}^{(n_1),...,(n_m)}(t)
\end{eqnarray}
which, for an arbitrary number of $m$ qubits, contains the
complete information on the tunnelling process as a function of
time via the counting variables $s\equiv\{s_i\}$ and the
conditional density matrices ${\rho}^{(n_1),(n_2),...}(t)$ for
$n_i$ tunnelling events (`jumps') to the drain $i$ after time $t$.
In matrix form, the EOM of the generating function follows from
the Liouville equation for the conditional density matrices and
reads $\dot{\textbf{G}}(s,t)=M(s)\textbf{G}(s,t)$ with formal
solution $\textbf{G}(s,t)=\exp[tM(s)]\textbf{G}(s,0)$. General
expectation values can be extracted from derivatives of
$\text{Tr}[\textbf{G}(s,\tau)]$ with respect
to the counting variables. 
In particular, the symmetrized noise correlation function
$S(\omega)_{i,j}\equiv \int_{-\infty}^{\infty} e^{i\omega \tau}
\langle \left\{\delta{I_i(t+\tau)},\delta{I_j(t)} \right\}\rangle $
between qubit $i$ and $j$ can then be written as a MacDonald formula
\cite{Mac48,Gur02}
\begin{eqnarray}\label{Mc_Donald}
\frac{S(\omega)_{i,j}}{2e^2\omega } &=& \int_0^{\infty}d\tau
\sin(\omega \tau) \frac{\partial}{\partial \tau}
\left \langle n_in_j -\frac{\tau^2\bar{I}_i\bar{I}_j}{e^2} 
 \right\rangle
\end{eqnarray}
where $\langle n_in_j\rangle
=\hat{D}_{ij}\text{Tr}[\textbf{G}(s,\tau)]|_{s=1}$ with the
differential operator $\hat{D}_{ij}\equiv
\partial_{s_i,s_j}+\delta_{ij}\partial_{s_i}$.
We simplify this expression following Flindt {\em et al.} \cite {Novotny_all} 
by introducing jump operators $L_i$ for qubit sources $i$ and
writing $\partial_\tau  \ex{n_in_j} = \text{Tr}[ L_i
\sum_{n_1,...,n_m} n_j
\rho^{(n_1),...,(n_m)}(\tau)]+(i\leftrightarrow j)$. This can be
further evaluated by Laplace transforming the EOM
$\partial_{t}\hat{G}(s,t)=(L_0 + \sum_i s_i L_i)\hat{G}(s,t)$
and taking derivatives in counting variables, giving
$\partial_{s_i}\tilde{G}(s,-i\omega)|_{s=1}=F_\omega
L_iF_\omega\rho_0$, where $F_\omega=(-i\omega-L)^{-1}$ and
$\rho_0$ is the steady state initial condition.  Using the
projections $F_\omega=-P/i\omega - R_\omega$, $R_\omega=Q(i\omega
+L)^{-1}Q$, ($P=\rho_0\otimes 1$, $Q=1-P$)  with  $P\rho_0=\rho_0$
and $Q\rho_0=0$ leads to
\begin{eqnarray}\label{Somega}
\frac{S(\omega)_{i,j}}{-2e^2}&=&\text{Re}\text{Tr}\left[\left (L_iR_\omega + \frac{\delta_{ij}}{2}\right) L_j\rho_0\right]+(i\leftrightarrow j).
\end{eqnarray}
In the zero frequency limit, we verify that the noise is
determined as usual \cite{BN03,Bel05} by the lowest eigenvalue
$\lambda_0(s)$ of the matrix $M(s)$, namely by the long-time
behaviour $G(s,t\to \infty) \propto \exp[t\lambda_0(s)]$ and
therefore
$S(0)_{i,j}={2e^2} \hat{D}_{ij} \lambda_0(s=1)$.

{\em Non-equilibrium  noise: results.--}
The currents through our two parallel charge qubits give rise to two diagonal and one off-diagonal
component in the tensor $S(\omega)_{i,j}$ of the noise spectrum.
In Fig.(\ref{fig_noise_spectrum}), we present results for the diagonal noise, i.e. the
 noise spectrum $S(\omega)_{1,1}=S(\omega)_{2,2}$ of the individual, interacting qubits.
This spectrum clearly displays resonances at the Bohr frequencies as given by the excitation
energies of the closed system. At $U=0$, there is one single resonance at $\omega=2T_c$ that splits up when $U$ is increased.
Similar to  light emission spectra in real molecules, frequency-dependent  shot-noise spectra thus provide
direct information about the correlated energy levels in artificial molecules.

The {\em cross-noise} spectrum exhibits a somewhat more
complicated resonance structure (inset of
Fig.(\ref{fig_noise_spectrum})).
More interesting is however the behaviour of the cross-correlation
Fano factor at zero frequency, $F(0)_{1,2}\equiv S(0)_{1,2}/2eI$,
which becomes positive as $U$ increases. This positive
cross-correlation is an indication of correlated emission of
electron pairs into different exit right leads \cite{positive}. As a
function of $U$, cf. in Fig.(\ref{fig_compare}), there is a strong
analogy between the cross-noise $F(0)_{1,2}$, its first derivative
$F'(0)_{1,2}$, and the non-equilibrium concurrence $C$, at least on
a qualitative level.
In particular, the non-analytic switching of $C$ with increasing $U$
from un-entangled to entangled states translates into a strongly
delayed (though smooth) onset of the increase in $F(0)_{1,2}$, and
the transition of $F'(0)_{1,2}$ from negative to positive. For
example, in the figure we see $F'(0)_{1,2}$ for $T_c=0.5$ becomes
positive around $U=1$ in agreement with the switching of $C$ at
$U\sim 2T_c^2/\Gamma_R=1$.  This analogy between noise and
entanglement so far holds on a qualitative level only.

\begin{figure}[t]
\begin{center}
\includegraphics[width=0.48\columnwidth]{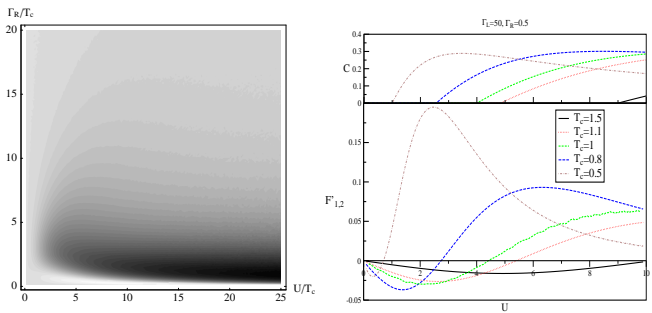}
\includegraphics[width=0.50\columnwidth,height=3.9cm]{fig4b.eps}

\caption{\textbf{Left:}  The cross-correlation zero-frequency
spectrum Fano Factor $F(0)_{1,2}=S(0)_{1,2}/2eI$.  The resemblence
to the concurrence, Fig. (\ref{fig_eq_noneq}),  is qualitative.
(White is minimum, $F_{1,2}=-0.12$, black is maximum,
$F_{1,2}=0.84$).  \textbf{Right:} The switching phenomenon in the
concurrence is more clearly seen, as well as a negative to positive
re-emergence in the first derivative of the noise around the same
point.} \label{fig_compare}
\end{center}
\end{figure}

Finally, we mention that we have not included the effect of dissipation in our calculations so far. Weak decoherence processes can in principle be easily incorporated through additional terms within the master equation. In \cite{AB04} it was shown how to use the resulting changes in the noise spectrum in order to extract, e.g., relaxation and
decoherence times $T_1$ and $T_2$. 
This can also be done for the interacting qubits discussed here.

{\em Conclusions}
We have shown how the entanglement of a
non-equilibrium double qubit differs from its 
thermal-equilibrium relative by exhibiting a $1/\Gamma$ switching
threshold for weak tunnelling rates $\Gamma$.  The cross-correlation
noise reflects this threshold at $\omega=0$ and shows  resonances at
the Bohr frequencies of the double qubit for finite $\omega$. Future
theoretical work may include clarifying this relationship and
checking the influence of decoherence on the correlated noise power
spectrum.

{\em Acknowledgements} We would like to acknowledge discussions and
suggestions from T. Novotn\'{y} and S. Debald. Work supported by the
Spanish MEC through grant MAT2005-07369-C03-03 and by the Japan
Society for the Promotion of Science Grant 17-05761.


\end{document}